%% file: main.tex
\documentclass[%
 preprint,
superscriptaddress,
nofootinbib,
 amsmath,amssymb,
]{revtex4-2}
\usepackage{amsmath}
\usepackage[section]{placeins}
\usepackage[utf8]{inputenc}
\usepackage[english]{babel}
\makeatletter\AtBeginDocument{\let\@elt\relax}\makeatother
\usepackage{float}
\usepackage{isotope}

\usepackage[dvipsnames]{xcolor}
\usepackage{cleveref}

\usepackage{soul}
\usepackage{graphicx}
\usepackage{dcolumn}
\usepackage{bm}

\usepackage{graphicx}
\usepackage{dcolumn}
\usepackage{bm}

\usepackage[draft, inline, nomargin]{fixme}
\fxsetup{theme=color, mode=multiuser}

\FXRegisterAuthor{bc}{1}{\color{red}BC}
\FXRegisterAuthor{ag}{2}{\color{blue}AG}
\FXRegisterAuthor{ph}{3}{\color{magenta}PH}
\definecolor{MH}{rgb}{0.0,0.6,9}

\newcommand\snowmass{\begin{center}\rule[-0.2in]{\hsize}{0.01in}\\\rule{\hsize}{0.01in}\\
\vskip 0.1in Submitted to the  Proceedings of the US Community Study\\ 
on the Future of Particle Physics (Snowmass 2021)\\ 
\rule{\hsize}{0.01in}\\\rule[+0.2in]{\hsize}{0.01in} \end{center}}

\newcommand{\JLAB}{\affiliation{Center for Advanced Studies of Accelerators,\\
 Jefferson Lab, Newport News, VA 23606}}

\newcommand{\NWU}{\affiliation{Physics \& Astronomy Department, Northwestern University, 
 Evanston, IL 60201-3112}}

\newcommand{\CNP}{\affiliation{Center for Neutrino Physics, Physics Department, Virginia Tech, Blacksburg, VA 24061}}

\newcommand{\FNAL}{\affiliation{Fermi National Accelerator Laboratory, Batavia, IL 60510}}

\newcommand{\CERN}{\affiliation{CERN, Esplande des Particules, 1211 Geneva 23, Switzerland}}
\newcommand{\ICL}{\affiliation{Imperial College London, Exhibition Road, London, SW7 2AZ, UK}}
\newcommand{\RAL}{\affiliation{STFC Rutherford Appleton Laboratory, Harwell Oxford, Didcot, OX11 0QX, UK}}

\newcommand{\PI}{\affiliation{Perimeter Institute for Theoretical Physics, Waterloo, ON N2J 2W9, Canada}}
\newcommand{\UMN}{\affiliation{School of Physics and Astronomy, University of Minnesota, Minneapolis, MN 55455, USA}}

\newcommand{\BNL}{\affiliation{Brookhaven National Laboratory, Upton, NY 11973}}
\begin{document}

\title{The Physics Case for a Neutrino Factory}

\author{Alex Bogacz}\JLAB
\author{Vedran Brdar}\NWU\FNAL
\author{Alan Bross}\FNAL
 \author{Andr\'e de Gouv\^ea}
 \email{degouvea@northwestern.edu}\NWU
\author{Jean-Pierre Delahaye}\CERN
\author{Patrick Huber}
\email{pahuber@vt.edu}\CNP

\author{Matheus Hostert}
\PI \UMN

\author{Kevin J. Kelly}\CERN
\author{K. Long}\ICL\RAL
\author{Mark Palmer}\BNL
\author{ J. Pasternak}\ICL\RAL

\author{Chris Rogers}\RAL


\author{Zahra Tabrizi} \NWU

\date{\today}%
\snowmass{}%




\maketitle

\tableofcontents
\newpage
\input{ask.tex}



\input{bsm.tex}

\input{detectors.tex}
\input{options.tex}

\input{synergies.tex}
\input{conclusion.tex}


\bibliographystyle{apsrev-title}
\bibliography{apssamp.bib}

\end{document}

%% file: ask.tex
\section{Executive Summary}

The discovery that $\theta_{13}$ is relatively large~\cite{DoubleChooz:2011ymz,DayaBay:2012fng,RENO:2012mkc} allowed the community to develop a promising experimental program based on scaling the beam power of conventional horn-focused neutrino beams along with the size of the associated detectors. These so-called super-beam experiments can reach the required statistical precision to discover (assuming it is there) and measure the parameters responsible for leptonic CP-invariance violation.  Two of these superbeam experiments are moving forward: the Deep Underground Neutrino Experiment (DUNE)~\cite{DUNE:2020ypp} and Hyper-Kamiokande (Hyper-K)~\cite{Hyper-KamiokandeProto-:2015xww}, and both are now under construction. Assuming the three-active-massive-neutrinos paradigm is correct, DUNE and Hyper-K will perform precision studies of the so-called atmospheric oscillation parameters, $\Delta m^2_{31}$ and $\theta_{23}$, and measure the leptonic CP-violation phase $\delta$, whereas the JUNO~\cite{JUNO:2021vlw} experiment, using reactor neutrinos, will determine $\theta_{12}$ and $\Delta m^2_{21}$, the so-called solar oscillation parameters. All three experiments have varying degrees of sensitivity to the neutrino mass ordering, but only DUNE  will perform precision studies of the MSW effect. In addition, they all also have a broad non-oscillation physics program ranging from geo-neutrinos to the detection of neutrinos from core-collapse supernovae.

Neutrino factories were first proposed around the time neutrino oscillations were first observed by Super-Kamiokande~\cite{Super-Kamiokande:1998kpq}. Neutrino factories provide a fundamentally different method for creating neutrino beams compared to pion decay-based horn-focused beams. Muon decay is a well understood process providing equal numbers of electron antineutrinos and muon neutrinos with precisely known energy spectra. Accelerating muons and putting them into a storage ring with long straight sections creates a well collimated neutrino beam~\cite{Geer:1997iz}; a muon antineutrino beam, along with an electron neutrino beam, is obtained by storing $\mu^+$ in the ring. The key advantages are a high luminosity, in particular also at high energies, both muon and electron flavor content, well known neutrino energy spectra and very well determined beam intensity. The challenge is production and acceleration of a sufficient number of muons to create a high-luminosity beam. This challenge is to a large degree shared with muon colliders and therefore, there is a strong synergy between these two programs. Given their excellent beam properties, neutrino factories are the ideal tool to study neutrino oscillations and thus a considerable effort to develop this concept ensued, which culminated in the International Design Study for the Neutrino Factory (IDS-NF)~\cite{IDS-NF:2011swj}. In the U.S., until the 2013 P5 report, the Muon Accelerator Program continued R\&D towards both a muon collider~\cite{Delahaye:2013jla} and a neutrino factory~\cite{Delahaye:2018yfq}.

A neutrino factory can exceed the precision, relative to conventional neutrino beams, with which the leptonic CP phase $\delta$ is measured and can potentially provide more precise measurements of $\Delta m^2_{31}$ and $\theta_{13}$. By exploiting the various available oscillation channels (including their CP conjugates), $\nu_\mu\rightarrow\nu_\mu$, $\nu_e\rightarrow\nu_\mu$, $\nu_\mu\rightarrow\nu_e$ and, depending on the stored-muon energies, the $\nu_\tau$ final states, a neutrino factory can over-constrain the three-active-massive-neutrinos paradigm in a way that is not accessible to any single ongoing or near future experiment, or combinations thereof. In particular, should the combined DUNE, Hyper-K and JUNO results reveal any surprises or anomalies, a neutrino factory would be the ideal tool to study these with a complementary approach and much smaller beam systematics.  

The sensitivity of neutrino factories to phenomena beyond the standard model (plus nonzero neutrino masses) has been studied in detail for the case of non-standard neutrino interactions~\cite{IDS-NF:2011swj}, especially for neutral current-like couplings. In this case, the access to potentially large matter effects at the neutrino factory allows unrivaled sensitivities. We argue that a neutrino factory, given its unique beam characteristics, should be a remarkable laboratory to explore a much wider range of new physics: new neutrino states, both heavy and light, different flavors of light dark matter, searches for lepton-flavor universality violation, and many others. One of the key advantages relative to conventional neutrino beams is the ability to study neutrino laboratory energies of order tens of GeV without compromising the beam luminosity. This opens, for example, the possibility for $\nu_\tau$ physics and charm production. Novel neutrino beams, both in energy and flavor content, invite the development of new fine-grained detectors which have  a high rate capability. Simple scaling from existing studies for conventional beams indicates a potentially very significant increase in sensitivity. 

There also is a renewed interest in  muon colliders \cite{AlAli:2021let,MCWP1,MCWP2} in the energy frontier, both as a Higgs-factory and as a high-energy-exploration enterprise, and there are large synergies at the machine level between those and neutrino factories. In particular, muon production, capture, and cooling, which are the most difficult challenges for a muon collider, would directly benefit a neutrino factory.

The rich physics opportunities offered by the very special beam characteristics at a neutrino factory and the need to prepare for a post-DUNE neutrino physics programs indicates that detailed studies of a neutrino factory complex, its physics reach and detectors are once again timely and needed.

In this Snowmass White Paper, we briefly highlight physics opportunities allowed by neutrino factories, see Section~\ref{sec:POatND}. The discussions here are meant to be illustrative and are neither exhaustive nor especially detailed. We discuss challenges and opportunities associated to neutrino detectors for the neutrino factory in Section~\ref{sec:DO}, and the current status and challenges associated with producing stored-muon neutrino beams in Section~\ref{sec:MO}. We especially highlight synergies between neutrino factories, muon colliders, and the FNAL accelerator complex in Section~\ref{sec:Synergies}. In Section~\ref{sec:conclusions}, we summarize the potential physics opportunities associated to neutrino factories in the post-DUNE and Hyper-K era, along with the steps required to ensure the community is prepared to seize them when the time is right.

%% file: bsm.tex

\section{Physics opportunities at the far site}

Neutrino factories are the ultimate tool to study three-flavor neutrino oscillation and can achieve precision measurements of $\sin^2\theta_{23}$, $\Delta m^2_{31}$, $\sin^22\theta_{13}$ and $\delta$, the leptonic CP phase. The also can over constrain the oscillation parameter by exploiting the $\nu_\mu$ and $\nu_e$ initial state flavors and  $\nu_\mu$ and $\nu_e$ final state flavors, and with the right detector even the $\nu_\tau$ final state. The achievable precision has been well documented in the literature, see for instance Ref.~\cite{IDS-NF:2011swj} and references therein, and is unmatched by DUNE~\cite{DUNE:2020jqi} and Hyper-K~\cite{Hyper-Kamiokande:2018ofw}. Once DUNE and Hyper-K results are available the community will need to evaluate if the additional precision offered by a neutrino factory is needed: if DUNE and Hyper-K results agree well with each other and three flavor oscillation the case would be much weaker than if surprises or discrepancies are found. In the latter case, the far site of a neutrino factory would be likely the best tool to address the open questions.

\begin{figure}[tp!]
  \begin{center}
    \includegraphics[width=0.9\textwidth]{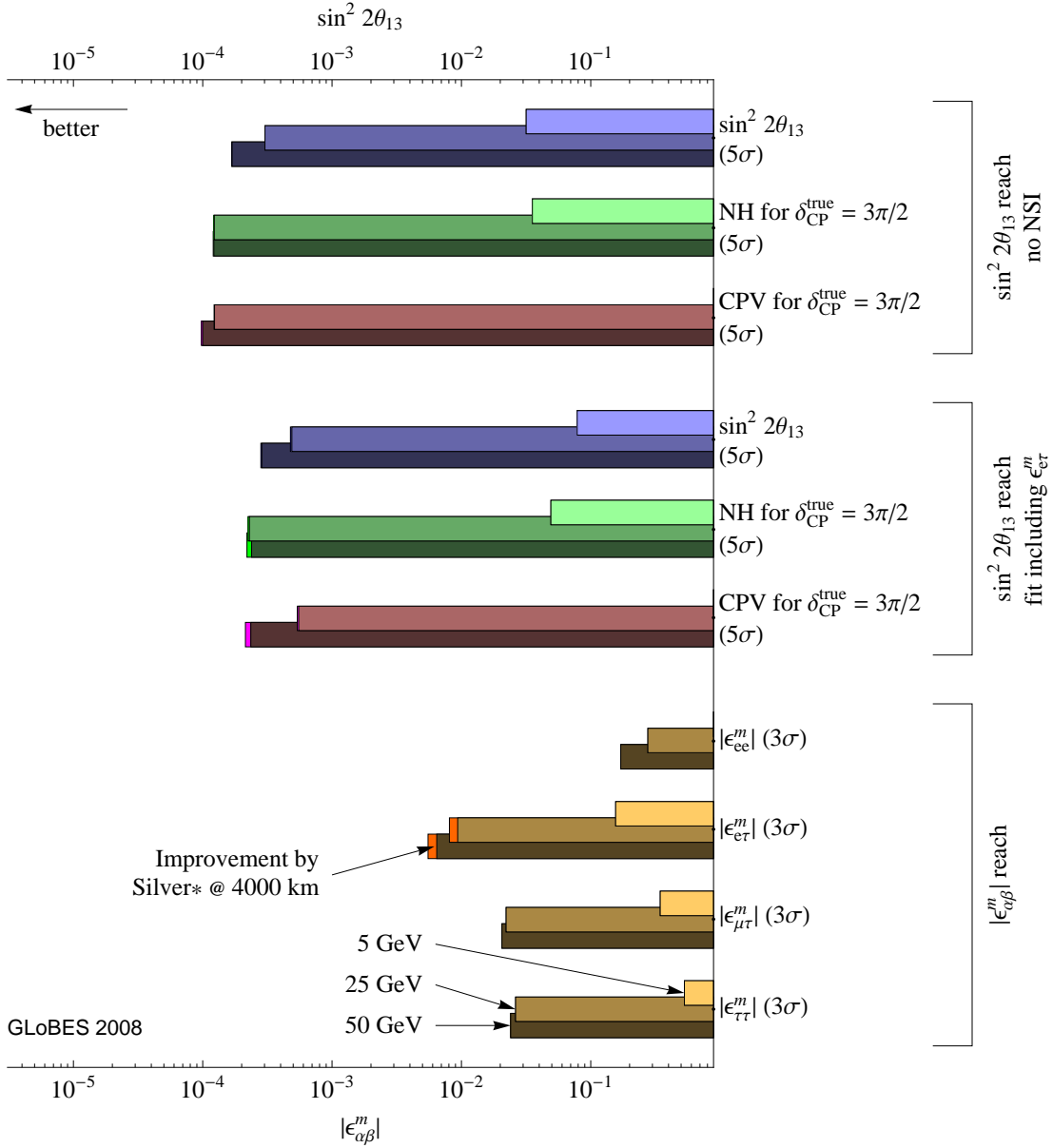}
  \end{center}
  \caption{\label{fig:summary}Summary for the optimization of a neutrino
    factory as a function of the muon energy. The dark bars represent $E_\mu=50 \, \mathrm{GeV}$, the medium light bars $E_\mu=25 \, \mathrm{GeV}$, and the light bars $E_\mu=5 \, \mathrm{GeV}$.  The upper group
of bars represents the standard optimization (in terms of the $\sin^22\theta_{13}$ reach), the middle group of bars represents the standard optimization (in terms of the $\sin^22\theta_{13}$ reach) including $\epsilon_{e\tau}$ marginalized over, and the lower group the non-standard optimization (in terms of the $| \epsilon_{\alpha \beta}^m|$ sensitivity).
Here the {\sf IDS-NF} setup is used with two baselines at $4 \, 000 \, \mathrm{km}$ and $7 \, 500 \, \mathrm{km}$. Both the sensitivities without silver channel, as well as with an advanced silver channel detector {\sf Silver*} are shown in all cases. As a benchmark point, $\sin^22\theta_{13}=0.001$ and $\delta=3 \pi/2$ has been chosen, as well as a true normal hierarchy. Figure and caption from Ref.~\cite{Kopp:2008ds}.
}
\end{figure}

With the same far detector complex, it also possible to conduct precision measurements of the matter effect and thus to constrain the existence of new neutrino interactions, aka non-standard interactions (NSI). In particular, neutral-current like NSI are very difficult to constrain directly by other means.
The sensitivity to NSI is shown in Fig.~\ref{fig:summary} and is clear that higher neutrino energies are preferred for this measurement.

Neutrino factories also offer excellent sensitivity to sterile neutrinos over a wide range of parameter space and in particular by studying anomalous $\nu_\tau$ appearance~\cite{Donini:2008wz}.  A neutrino factory allows to disentangle the magnitudes of $|U_{e4}|, |U_{\mu 4}|$ and $|U_{\tau 4}|$. The far site sensitivity arises in the same way as DUNE is sensitive~\cite{Dutta:2016glq} to sterile neutrinos at the far site: the sterile oscillation is averaged out but still leads to an effect in oscillations by modifying the amplitude of the oscillation. Given the richer flavor content of the beam, higher beam energy and luminosity, the neutrino factory sensitivies exceed those DUNE or Hyper-K significantly.

\section{Physics Opportunities at the near site}
\label{sec:POatND}

As will be described below, the near detector at the Neutrino Factory will have a rich program searching for new physics as well as having a long list of neutrino interaction topics that can be explored with unprecedented precision.  In order to take full advantage of the opportunities at the near site, the detector requirements extend far beyond what is needed for the 3-flavor oscillation search.  Options for the near detectors are discussed in Section~\ref{sec:DONS}, but we list some of the overarching performance requirements here:
\begin{itemize}
\item Highly segmented detectors capable of precision operation at high event rate.  Detectors with inherent 3D tracking (or very precise timing) capability over 4$\pi$ are required.
\item Excellent muon and electron ID capability.
\item Excellent energy resolution. 
\item A magnetized detector for charge identification.  In addition, reconstruction via spectrometry can be applied to event reconstruction as opposed to being done via calorimetry.  This is particularly important for high-energy ($E_\nu \ge$ 10 GeV) neutrino interactions where the outgoing muon's momentum must be measured via spectrometry.
\item Excellent particle ID, ie, p/$\pi$/K separation at momenta from a few hundred MeV/c to a few GeV/c.
\item Neutron detection capability (with energy determination).
\item A variety of nuclear targets to measure cross-sections as a function of the nuclear target mass number A.
\item Micron-scale resolution for charm and tau identification or the capability to tag charm and taus in the final state via kinematics.
\end{itemize}

With these requirements in mind, we discuss the prospects of measurements that can be performed at the neutrino factory near detector.

\subsection{Precision in Cross Section Measurements} \label{sec:HEprecision}

A neutrino factory with energies above several tens of GeV would provide a new precision probe of the weak interactions and of the nucleon structure. The detailed knowledge of the neutrino flux, both in overall normalization and shape, coupled with a high-granularity detector allows for the precise determination of exclusive neutrino-nucleus cross-sections. At the highest energies, deep-inelastic-scattering (DIS) dominates the neutrino cross section with nucleons and one can study the nucleon structure at low Bjorken $x$ and high $Q^2$. In the past, the precision physics program in neutrino DIS relied on ratios of event rates in order to mitigate the large neutrino flux uncertainties, which was plagued by uncertainties from the subtraction of $\nu_e$ contamination backgrounds. For a neutrino factory, all components of the beam are well-known and the extraction of the neutrino cross sections can be performed directly and with much greater precision. 

Contrary to standard neutrino beams, a muon-decay neutrino beam has an equal number of $\overline{\nu}_\mu(\nu_\mu)$ and $\nu_e (\overline{\nu}_e)$ in $\mu^+(\mu^-)$-storage mode. While this allows to study $\nu_e(\overline{\nu}_e)-N$ scattering  with large statistics, it may also present new backgrounds to NC measurements. A high-purity measurement of the neutral current rate at a high-resolution detector would require excellent $\nu_e(\overline{\nu}_e)-N$ CC and $\nu-N$ NC discrimination, since electrons can mimic NC interaction more often than muons. Since the low density tracker at the Neutrino Factory can efficiently reconstruct the electron tracks, the $\nu_e$ CC interactions can be identified on an event-by-event basis, reducing the contamination to a negligible level. Similarly, uncertainties related to the location of the interaction vertex, noise, counter efficiency etc. are removed by the higher resolution and by the different analysis selection.

\subsection{Standard Candles} At conventional neutrino facilities where the flux is poorly known, measurements of well-known scattering processes, i.e. ``standard candles", can help reduce flux uncertainties. Three main processes have been often discussed: 
\begin{itemize}
    \item low-$\nu$ neutrino-nucleus scattering for the flux normalization and shape,
    \item coherent meson production for the flux shape,
    \item elastic neutrino-electron scattering for the flux normalization,.
\end{itemize}
Each of these processes can be measurement with a precision as good as the flux and fiducial volume uncertainties at a neutrino factory, which would be well below the percent level. 

Low-$\nu$ interactions are neutrino-nucleus CC processes with small energy transfer to the nucleus, $\nu = E_{\rm had} = E_\nu - E_\ell$. In other words, the inelasticity $y = \nu/E_\nu$ is small and so is the recoil hadronic energy, leaving the nuclear target mostly undisturbed. In this regime, the neutrino cross section only depends on the structure function $\mathcal{F}_2$ and is a linear function of energy. That is,
\begin{equation}
    \frac{d\sigma}{d y} \xrightarrow{y \ll 1} \frac{G_F^2 M E_\nu}{\pi} \int \mathcal{F}_2(x) \,d x = ({\rm  Constant}) \times E_\nu,
\end{equation}
where $x = Q^2/2 M \nu$. Therefore, a measurement of the the low-$\nu$, or equivalently, low-$y$, cross section across the range of values of $E_\nu \approx E_\ell$ in the experiment can determine the normalization and shape of the flux. At a neutrino factory, a low-density detector with low-thresholds would allow to probe the very small $y$ piece of the cross section with minimal finite-$y$ corrections. This is to be constrasted with previous measurements of the low-$\nu$ cross section by MINOS and NOMAD, with precision of $\gtrsim 3~\%$~\cite{Bodek:2012uu}.

Coherent meson production also leaves the nucleus undisturbed. In this case, for example coherent $\ell^\pm \pi^\mp$ production allows one to measure the incoming neutrino energy as $E_\nu \approx E_\ell + E_\pi$, as well as its direction, since $\vec{p}_\nu \approx \vec{p}_\ell + \vec{p}_\pi$. A precise measurement of the total value and shape of this cross section would allow other experiments to determine the shape and normalization of their fluxes more precisely. Once more, the low-threshold advantage of a HPgTPC detector would allow for a greater reduction of backgrounds from incoherent processes.

\subsection{Precision in $\sin^2{\theta_{\rm W}}$ Measurement}

\paragraph{Revisiting NuTeV} 

To reduce the impact of flux systematic uncertainties on precision DIS measurements, NuTeV made use of a Paschos-Wolfenstein ratio~\cite{Paschos:1972kj}. The observation is that, up to small corrections, in a pure $\nu_\mu$ or $\overline{\nu}_\mu$ beam, the ratios between neutral- and charged-current scattering rates on an isoscalar target follow 
\begin{equation}\label{eq:Rratio}
R^{\nu} = \frac{\sigma(\nu_{\mu}N\rightarrow\nu_{\mu}X)}
                 {\sigma(\nu_{\mu}N\rightarrow\mu^-X)}  
\simeq \frac{1}{2}-\sin^2\theta_{\rm W}+\frac{5}{9}(1+r)\sin^4\theta_{\rm W},
\end{equation}
and
\begin{equation}
R^{\overline{\nu}} = \frac{\sigma(\overline{\nu}_{\mu}N\rightarrow\overline{\nu}_{\mu}X)}
                 {\sigma(\overline{\nu}_{\mu}N\rightarrow\mu^+X)}  
\simeq \frac{1}{2}-\sin^2\theta_{\rm W}+\frac{5}{9}\left(1+ \frac{1}{r}\right)\sin^4\theta_{\rm W},
\end{equation}
where $r \equiv \sigma({\overline \nu}_{\mu}N\rightarrow\mu^+X)/
\sigma(\nu_{\mu}N\rightarrow\mu^-X) \simeq \frac{1}{2}$. In reality, the relation above is extended to include second and third generation quarks, differences between $\mu^+$ and $\mu^-$ reconstruction, radiative corrections, contributions from non-isoscalar targets, and corrections from contamination in the neutrino beam. The even more powerful ratio,
\begin{equation}
    R^- = \frac{R^\nu - r R^{\overline{\nu}}}{1 - r} = \frac{1}{2} - \sin^2{\theta_{\rm W}},
\end{equation}
is independent of heavy quark corrections and allowed NuTeV to reach a precision on the weak-mixing angle $\theta_{\rm W}$ of order $1\%$. The largest experimental systematic uncertainty in NuTeV is related to the subtraction of the $\nu_e$ CC contamination from the NC sample. However, the NuTeV deviates from the measurement done by LEP at the $3\sigma$ level. Therefore, a precise measurement of the weak mixing angle by neutrino scattering data, specially at the same energy range of NuTeV helps in understanding the situation. 

As discussed before, at a neutrino factory, the electron and muon flavors are comparable and new techniques would be required.

\paragraph{Probing the Running of $\sin^2{\theta_{\rm W}}$.} 
A neutrino factory can measure the weak mixing angle through different channels: i) Deep inelastic scattering of neutrinos off quarks (see above). This has a center of mass energy of $s=2xE_\nu m_N$, where $x$ is the Bjorken parameter and $m_N$ is the nucleon mass; ii) Elastic scattering of neutrinos off electrons, with $s=2E_\nu m_e$; iii) Elastic scattering of neutrinos off protons, with  $s=2E_\nu m_N$. All these channels have considerably different momentum transfer scales, and hence can give the information on the running of the weak mixing angle. As explained above, for DIS, the measurement could be performed using the Paschos-Wolfenstein ratio, as done at NuTeV.   For the $\nu-e$ scattering channel the experiment can look at the ratio of the $\bar\nu_\mu-e$ over $\nu_\mu-e$ channel, where the systematic uncertainties which are related to the  electron identification and selection cancel out. However, this channel is limited by lower statistics. DUNE expects percent-level uncertaity on this measurement at $q\simeq 30$~MeV, using approximately $60,000$ expected $\nu-e$ scattering events in 7 years of data collection~\cite{deGouvea:2019wav}. However, the measurement at DUNE is limited by higher flux uncertainties. On the other hand, at the few GeV energy range of DUNE the uncertainties on the neutrino-nucleus cross section makes in very challenging to use this channel for measuring the weak mixing angle. The higher energy range at the neutrino factory  as well as lower systematic uncertainties on the neutrino fluxes make the near detectors ideal for weak mixing angle measurement. 


\subsection{Light Sterile Neutrinos}\label{sec:Steriles}
With a well-characterized beam of $\nu_e$ at a neutrino factory, we can perform precision searches for $\nu_e \to \nu_\mu$ and $\nu_e \to \nu_\tau$ appearance at short-baselines, where this new oscillation can be driven by a fourth, sterile neutrino. These new oscillation probabilities are (for $\alpha = \mu,\ \tau$)
\begin{equation}
    P\left(\nu_e \to \nu_\alpha\right) = \sin^2\left(2\theta_{e\alpha}\right) \sin^2\left(\frac{\Delta m_{41}^2 L}{4E_\nu}\right),
\end{equation}
where $\sin^2\left(2\theta_{\alpha\beta}\right)$ is an effective mixing angle depending on the extended $4\times4$ neutrino mixing matrix, $\Delta m_{41}^2 \equiv m_4^2 - m_1^2$ is the new mass-squared splitting, $L$ is the distance between neutrino production and detection, and $E_\nu$ is its energy. When using natural units, the argument of the oscillatory term becomes
\begin{equation}
    \frac{\Delta m_{41}^2 L}{4E_\nu} \longrightarrow 1.267 \left(\frac{\Delta m_{41}^2}{\mathrm{eV}^2}\right) \left(\frac{L}{\mathrm{km}}\right) \left(\frac{\mathrm{GeV}}{E_\nu}\right).
\end{equation}
Given the multiple detectors that we are considering, this oscillation probability can be measured at multiple values of $L$ for common $E_\nu$, bolstering the searches for anomalous appearance of $\nu_\mu$ or $\nu_\tau$ arising in the $\nu_e$ beam. Furthermore, measurements of the $\nu_e$ flux at both detector locations allows for reducing systematic uncertainties.

Given the neutrino energy and the detector positions, we may determine the range of $\Delta m_{41}^2$ that the neutrino factory can probe. A previous neutrino factory proposal studied the capability of a magnetized iron-scintillator far detector, nuSTORM~\cite{nuSTORM:2014phr}, with mass 1.3 kt, located 2 km from the end of the muon storage ring (along with a $\mathcal{O}(100) ton$ near detector 50 m from the end of the storage ring). In Ref.~\cite{nuSTORM:2014phr}, a ten-year exposure, corresponding to $10^{21}$ protons on target yielding muons with momentum $3.8$ GeV ($\pm 10\%$) was considered. The sensitivity to $\sin^2\left(2\theta_{e\mu}\right)$ in regions of interest~\cite{PhysRevLett.77.3082,MiniBooNE:2010idf,MiniBooNE:2020pnu} for $\Delta m_{41}^2 \approx$ 1 eV$^2$ is impressive, reaching angles on the order of $10^{-4}$ at 99\% CL, even with (realistic) 1\% systematic uncertainties. This is in contrast with the capability of the joint Fermilab SBN detectors, with an expected sensitivity of $\sin^2\left(2\theta_{e\mu}\right) \approx 5\times 10^{-4}$ in the next several years~\cite{Machado:2019oxb}. In the future (by roughly 2040), DUNE expects sensitivity at the level of $\sin^2\left(2\theta_{e\mu}\right) \approx 10^{-6}$ with its near detector complex~\cite{DUNE:2020fgq}. However, such a strong constraint depends significantly on the level of uncertainties assumed, and the neutrino flux uncertainties from a proton-beam-source such as LBNF would be naturally larger than at a neutrino factory.

If we consider a higher muon-beam energy (e.g. one or two orders of magnitude higher-energy neutrinos) than what was assumed for nuSTORM, then we would be sensitive to larger values of $\Delta m_{41}^2$, in the range of ${\sim}10-100$ eV$^2$. Measurements in such a range would complement those coming from direct neutrino-mass-measurement experiments such as KATRIN~\cite{KATRIN:2020dpx,KATRIN:2022ith}. We note that KATRIN, and other beta-decay experiments, are sensitive to the $3+1$ sterile neutrino mixing angle $\sin^2\left(2\theta_{ee}\right) \equiv 4\left|U_{e4}\right|^2\left(1 - \left|U_{e4}\right|^2\right)$, contrasted against neutrino-factory searches for $\sin^2\left(2\theta_{e\mu}\right) \equiv 4\left|U_{e4}\right|^2 \left|U_{\mu 4}\right|^2$. This complementarity would be extremely valuable in the event of a discovery in one or both of the search environments.

\subsection{Lepton Number Violation}\label{sec:nunubar}

 Neutrino factory beams are very pure, that is only muon neutrinos (antineutrinos) are present in the beam for stored $\mu^-$ ($\mu^+$), and similarly, only electron antineutrinos (neutrinos) would be present. In this environment, a search for neutrino-antineutrino conversion would have great sensitivity due to the magnetized nature of the detector. This is especially true for the $\nu_\mu$ ($\overline{\nu}_\mu$) flux, since $\mu^+$ and $\mu^-$ do not shower like $e^+$ and $e^-$ pairs do. 

A study for nuSTORM with a magnetized ($B \sim 2$~T) iron detector found a rejection of wrong-sign muons to be $5\times 10^{-5}$, with a signal efficiency $0.16$. Depending on the size of the magnetic field, these numbers are expected to be significantly improved at a HPgTPC.

Due to the high energies of a neutrino factory, processes exclusive to neutrinos in the beam can be exploited. For instance, inverse muon decay,
\begin{equation}
    \nu_\mu e^- \to \nu_e  \mu^- \text{  and   } \overline{\nu}_e e^- \to \overline{\nu}_\mu \mu^-,
\end{equation}
with a threshold of $E_\nu \gtrsim 11$~GeV can be identified due to the forwardness of the muon and would only take place in $\mu^-$ mode, but not in $\mu^+$. Any deviation from this pattern would indicate new physics. This type of search would be sensitive to effects of lepton number violation (LNV) in muon decays from high-scale as well as low-scale extensions of the SM. 

\paragraph{LNV from heavy new physics} At the effective operator level, one can consider
\begin{equation}
    \frac{1}{\Lambda^5}(\overline{L_\mu} H e_R)(\overline{L^c_e} L_\alpha HH) \to \frac{v_{\rm EW}^3}{\Lambda^5} (\overline{\mu}_L e_R)(\overline{\nu^c_e} \nu_\alpha),
\end{equation}
and the corresponding right-handed muon Yukawa contraction, $\mu_R L_e H^\dagger$, contribute to an effective neutrino-antineutrino conversion in muon beams. Muons in the storage ring would undergo the following LNV decays,
\begin{equation}
    \mathcal{B}(\mu^+ \to e^+ \overline{\nu}_e \overline{\nu}_\alpha) \sim \left(\frac{v^3_{\rm EW}}{4 \sqrt{2}G_F \Lambda^5} \right)^2 \sim 0.1\%\, \left(\frac{370 \text{GeV}}{\Lambda}\right)^{10},
\end{equation}
and analogously for $\mu^-$. These branching ratios are at the level of the best constraints coming from measurements of the Michel parameters of muon decays~\cite{Armbruster:2003pq}. These models have been discussed in the context of the LSND anomaly in Refs.~\cite{Babu:2002ica,Babu:2016fdt}, where the authors also provide feasible UV completions. The unique capabilities of a neutrino factory would allow to study $\nu_e \to \overline{\nu_e}$ conversions that are otherwise hard to do in conventional accelerator beams due to the large wrong-sign-neutrino contamination. 

\paragraph{LNV from decaying-sterile neutrinos} Low-scale extensions of the SM can also lead to effective LNV. The simplest extension of the $3+1$ oscillation model discussed in Sec.~\ref{sec:Steriles} by a $\nu_s$-philic singlet scalar $\varphi$ can lead to decays in the neutrino beam that cascade down to neutrino-antineutrino paris. Through neutrino mixing with the muon or electron flavor, $\nu_4$ states can be produced in $\mu$ decays, which then decay in flight via $\nu_4 \to \nu \varphi$, or even $\nu_4 \to \overline{\nu} \varphi$ in the case of Majorana neutrinos. The latter decays would lead to an effective neutrino-antineutrino conversion, with the wrong-sign neutrinos having a bit less energy than the average particles in the beam. Even for Dirac neutrinos, if $\varphi$ is massive and decays to neutrino-antineutrino pairs via $\varphi \to \nu \overline{\nu}$, an excess of wrong-sign neutrinos would be noticeable. Constraints on these scenarios come primarily from searches for solar antineutrinos(see, e.g., Ref~\cite{Hostert:2020oui}). However, solar neutrino experiments are not sensitive to models with $|U_{e4}| \ll 10^{-3}$ or to models where $m_4$ or $m_{\varphi}$ are greater than $16$~MeV, since in that case the particles would not be produced in the decays of $^{8}B$ in the Sun.

\paragraph{LNV from lepton-number-charged scalars} Finally, light scalars carrying lepton-number can lead to effective $\nu \to \overline{\nu}$ transitions in two scenarios: if they are produced in 
\begin{equation}
    \mu^+ \to e^+ \overline{\nu} \, \overline{\nu} \, \varphi_{L=2}
\end{equation}
or in the scattering of neutrinos with matter, 
\begin{equation}
    \nu p^+ \to \ell^+ \, n \, \varphi_{L=2}.
\end{equation}
Neutrino experiments provide leading constraints on this scenario for regions of parameter space where $m_\varphi \gtrsim 100$ MeV, with future prospects from DUNE~\cite{Berryman:2018ogk,Kelly:2019wow} and the Forward Physics Facility~\cite{Kelly:2021mcd} presenting particularly strong capabilities. For a review of this type of scalar, see Ref.~\cite{Berryman:2022hds}. Searches for this type of anomalous $\overline\nu$ appearance in the beam may exceed current and future sensitivities.

\newcommand{\zt}[1]{{\bf\color{teal}(ZT: #1)}}

\subsection{NSI and New Physics at the Multi-TeV scale} \label{sec:EFT}
A near detector at the neutrino factory with the ability to measure the charm cross-section can measure $\tau$'s as well. Hence, even though the neutrino fluxes at the near site will purely consist of electron and muon neutrinos, non-standard four-fermion interaction of neutrinos with charged leptons and quarks can induce non-zero number of $\nu_\tau$ at the detector. These charged current (CC) four-fermion interactions can modify the production  and detection of neutrinos and we can study them using a systematic Effective Field Theory (EFT) approach (see e.g. Refs.~\cite{Falkowski:2019xoe,Falkowski:2019kfn,Falkowski:2021bkq} for EFT formalism at neutrino experiments). After obtaining constraints on the parameters of low-energy EFT relevant for the energy range of the neutrino factories, we can match them to the Wilson coefficients of the so called SM EFT (SMEFT) Lagrangian and indirectly probe new physics at energies much higher than the electroweak scale. 

In the presence of non-standard interactions in the muon decay, $\tau$ neutrinos can be produced in the beam. On the other hand, as a result of non-SM interactions on the detection side through the deep inelastic scattering of neutrinos on nucleons (DIS), extra number of $\nu_\tau$ can be detected. This results into a non-zero rate of $\nu_\tau$ events at the detector given by~\cite{Falkowski:2019kfn,Falkowski:2021bkq}:
\begin{equation}
    \frac{dN^\tau}{dE_\nu}\simeq  N_T \sigma_\tau^{\rm{SM}}\sum_{\alpha=e,\mu}\phi_\alpha^{\rm{SM}} \sum_X\Big\{p_{XX,\alpha}^{\mu-{\rm{decay}}}|[\epsilon_X^{\mu-{\rm{decay}}}]_{\alpha\tau}|^2+d_{XX,\tau}^{{\rm{DIS}}}|[\epsilon_X^{{\rm{DIS}}}]_{\tau\alpha}|^2\Big\}\,,
\end{equation}
where $N_T$ is the total number of target particles at the near detector, $\sigma_\tau^{\rm{SM}}$ is the SM DIS cross section of $\tau$ neutrinos, $\phi_\alpha^{\rm{SM}}$ is the SM flux of electron and muon neutrinos in the beam, and $X=L,R,S,P,T$ denotes to new left(right) handed, (pseudo)scalar and tensor interactions. The new physics effects at the production and detection side are shown by coefficients $p_{XX,\alpha}$ and $d_{XX,\tau}$, which depend on the neutrino energy and can be found in Ref.s~\cite{Falkowski:2019kfn,Falkowski:2021bkq}. For the muon decay (DIS), these coefficients are dominant for right handed (right handed and tensor) interactions and are of the order of 1. All in all, with the expectation of $10^9$ SM $\nu_{e,\mu}$ events at the near detector of a neutrino factory, we can expect to get constraints of $[\epsilon_{R}]_{\alpha\tau}\lesssim 10^{-4}$ from muon decay, and $[\epsilon_{R,T}]_{\tau\alpha}\lesssim 10^{-4}$ from DIS, translating in the new physics reach of $\Lambda\equiv v/\sqrt{\epsilon}\sim$ tens of TeV. Depending on the specific quarks in the four-fermion interactions, these constraints are one to two orders of magnitude better than relevant constraints from either low energy pion decay of high energy LHC results (see Tables 3-4 of \cite{Falkowski:2021bkq} for comparision).  

\subsection{Light Dark Sectors, Dark Matter}  \label{sec:LightMediators}

\newcommand{\vb}[1]{{\bf\color{RubineRed}(VB: #1)}}

While neutrino experiments have already delivered groundbreaking 
results in terms of measurements of leptonic mixing angles and neutrino mass
squared differences, they have also left us with a couple of 
puzzles, namely hints for neutrino oscillations at very short baselines. Here, 
acceleration-based neutrino experiments LSND \cite{PhysRevLett.77.3082} and MiniBooNE \cite{MiniBooNE:2010idf,MiniBooNE:2020pnu} stand out.\footnote{For completeness let us also mention reactor \cite{Mention:2011rk,PhysRevC.84.024617} and gallium anomalies \cite{Giunti:2010zu} at MeV-scale energies which appear less relevant in the context of this document; in particular, we should also stress that in lightof recent reactor flux measurements, reactor anomaly seems to be fading away \cite{Giunti:2021kab}} What is common between LSND, MiniBooNE and potential neutrino factory is the production of neutrinos: accelerated protons hitting a target or a beam dump which yields production of mesons that decay to neutrinos. This suggests that, in general, new physics scenarios that explain LSND and/or MiniBooNE can be tested at neutrino factories. Further, given that the typical neutrino energy from neutrino factory is 1-2 orders of magnitude larger, various BSM realizations can be probed even at the higher mass scales. The vanilla explanation for LSND and MiniBooNE boils down to the existence of eV-scale sterile neutrino that mixes with both muon and electron neutrinos, thus allowing for efficient transition between $\nu_\mu$ and $\nu_e$ at short baselines; this in turn yields an enhancement in the number of $\nu_e$ at the detector, explaining the observed excess of electron-like events. Sterile neutrino explanation is however disfavored from muon disappearance searches performed by IceCube \cite{IceCube:2020phf} and MINOS \cite{MINOS:2017cae}. To this end, there has recently been a number of alternative explanations to explain anomalies, primarily MiniBooNE, see e.g. \cite{Fischer:2019fbw, Gninenko:2009ks, Bertuzzo:2018itn, Dentler:2019dhz, Ballett:2018ynz,
deGouvea:2019qre, Abdallah:2020biq, Dutta:2020scq, Datta:2020auq, Abdallah:2020vgg,
Abdullahi:2020nyr, Brdar:2020tle,Abdallah:2020vgg}. Let us illustrate only one of the scenarios, e.g. the one introduced in \cite{Bertuzzo:2018itn}. There, the excess of electrons in the detector is explained by the upscattering of (primarily) muon neutrinos into right-handed ``dark neutrinos,'' where the latter undergo a 2-step decay; dark neutrinos first decay into 
a neutrino and an on-shell $U(1)_X$ gauge mediator which in turn yields an $e^+ e^-$ pair that 
can appear as a single shower event in MiniBooNE. This model was previously tested using MINERvA data \cite{Arguelles:2018mtc}, but given an existing near detector technology that we envision to be placed in the vicinity of a neutrino factory, such facility would also be highly sensitive. 
For completeness, let us stress that while the extended dark sectors (such as the one discussed above) can be tested, neutrino factories are also sensitive to minimal 
extensions of the Standard Model, e.g. light vector bosons \cite{Ballett:2019xoj} or
 heavy right-handed neutrinos.

Above, we have chiefly focused on neutrino-induced new physics events. Let us state that proton collisions with the target also yield a lot of photons and neutral mesons. Here, the former can produce hidden electrically neutral states in scattering while the latter can partially decay into hidden states. This was recently explored in the context of axions and axion-like particles \cite{Kelly:2020dda,Brdar:2020dpr} for DUNE and a neutrino factory is also sensitive to these scenarios.

One additional attractive scenario that can be tested from decays of neutral particles, primarily mesons, is the light, GeV-scale, dark matter model. Mesons are envisioned to partially decay into dark photons, and this is realizable provided dark photons have a nonzero kinetic mixing with the SM photons. Dark photons are produced on-shell or off-shell depending on their mass relative to the decaying meson. The coupling of dark photon to dark scalar (or fermion) states ensures that it promptly decays and the daughter particles then travel to the detector where they can scatter on electrons; see \cite{Kelly:2020dda,Breitbach:2021gvv} in the context of the DUNE near detector. Such dark scalars/fermions can satisfy, due to the thermal freeze-out mechanism, the relic abundance of dark matter in the universe in agreement with observations~\cite{Pospelov:2007mp}.  

The neutrino-induced scatterings  present a background for testing such dark matter models. This background can be chiefly eliminated if detector is placed off-axis. In such case, background from charged pion decays is suppressed because charged states are focused in the forward direction. An ideal solution at neutrino factory site would be DUNE-PRISM-like \cite{Kelly:2020dda} detector that is movable in the plane perpendicular to the beam. The employment of such detector and measurements at various sites would not only maximize background discrimination for this dark matter model, but would also allow the flexibility in light of various scenarios that can be scrutinized. Namely, for testing models that explain MiniBooNE, such a detector would need to be placed on axis. We point out that movable detector is not crucial for BSM goals at neutrino factories, but it would certainly aid in maximizing its full physics potential. Also the sophisticated front end of a neutrino factory may allow to disentangle the neutrino beam and meson flight directions at the target station, which is an interesting option to eliminate neutrino backgrounds.

\subsection{Decay in flight of new particles.}\label{sec:DIFsearches}

 Using a low-density detector made of Helium bags, NuTeV observed three anomalous $\mu^+\mu^-$ events in a search for the decay in flight of HNLs. The search was performed using a slightly off-axis (${\sim}$few mrad) location in a zero-background search. If the rate were due to muon production in neutrino interactions in the Helium bags, it would have exceeded the predicted Monte Carlo rate by two orders of magnitude. In a neutrino factory, this measurement could be performed in a HPgTPC, achieving a similar suppression of the neutrino interaction backgrounds as the Helium bags at NuTeV, and it would benefit from the use of a magnetic field. An off-axis detector location may benefit from additional suppression of neutrino interaction backgrounds.

%% file: detectors.tex
\section{Detector options}
\label{sec:DO}

A unique strength of the Neutrino Factory is the ability to provide high-energy neutrino beams without a sacrifice in neutrino intensity.  Although there is a wide ranging physics program at the NF, its reach at high-energy is unique among man-made neutrino sources and the detectors at both the near and far sites must be able to take advantage of this capability.  Since the NF is a facility for ``precision" physics, the detectors must be high resolution and capable of accurately tagging exclusive final states.
\subsection{Far Site}
\label{sec:DOFS}
Due to their required large size, options for the Far detectors are limited.  Certainly large water Cerenkov, LAr and magnetized iron detectors are options, but of these currently available technologies only the magnetized-iron detector is capable of fully containing the events that would be present at a high-energy (stored E$_\mu$ = 50 GeV for example) neutrino source from a NF and measure the charge of the final-state muon.  The maximum energy of a final-state muon in a $\nu_\mu$CC events that can be analyzed via range is $\simeq$ 12 GeV (in a very limited volume) for the large water Cerenkov detector of HyperK~\cite{Wilson:2021uwb} and $\simeq$ 4 GeV for DUNE's~\cite{Palomares:2021vba} LAr detectors at the far site.   
\subsection{Near Site}
\label{sec:DONS}
Many of the detector concepts now incorporated in the upgraded T2K near detector~\cite{Blanchet:2021pqb} and those being developed for DUNE~\cite{DUNE:2021tad} are appropriate for near detectors at a NF.  These concepts include:
\begin{enumerate}
    \item Highly-segmented tracking scintillator detector (SuperFGD)
    \item Pixelated LAr detector
    \item Magnetized high-pressure gaseous Ar TPC (HPgTPC)
    \item Straw-Tube trackers (STT) with thin targets
\end{enumerate}
Magnetization of all these detectors is under consideration.  The SuperFGD for T2K is a magnetized detector as will be the STT for DUNE.  The HPgTPC is by design a magnetized detector.  Although magnetization concepts for a pixelated LAr have been developed, the high cost for the magnet system presents obstacles to its utilization, although R\&D on high-temperature superconductor and cable may make this option affordable.  Without magnetization a LAr detector at the near site for a NF will need a muon catcher, since the LAr will not be large enough to fully contain most muons.

\begin{figure}[h]
\label{fig:ND-GAr}
\centering
\includegraphics[width=0.7\textwidth]{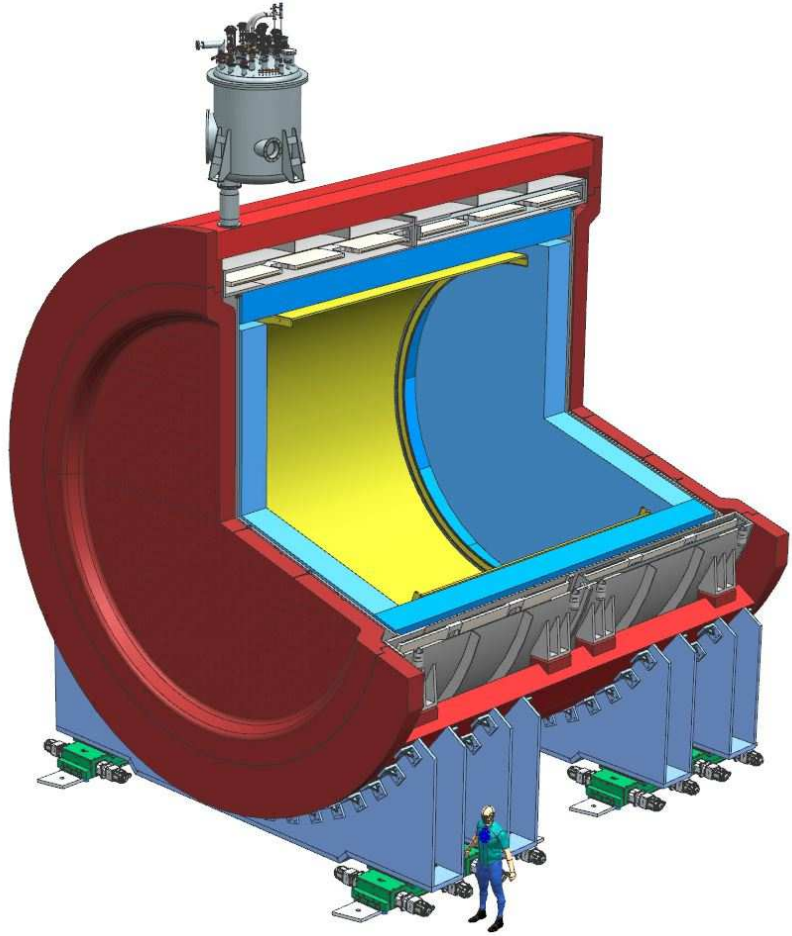}
\caption{Schematic of the ND-GAr concept.  From inside out, Yellow:HPgTPC, Blue:ECAL, Grey: Superconducting solenoid in its cryostat and Red: Return Iron.}
\end{figure}

Although the concept of magnetization in neutrino detectors is not new, the application of a collider-detector design for neutrino physics is.  One such example is the high-pressure gas TPC (HPgTPC) detector concept (called ND-GAr) for the DUNE near detector complex~\cite{Bersani:2021blp}.  An overview of the detector is shown in Figure~\ref{fig:ND-GAr}.
ND-GAr is a large detector with a magnetic volume that is approximately 7m in diameter and 7.5m long (both the HPgTPC and the ECAL are in the magnetic volume).  The solenoid magnet and the return iron provide pressure containment.  This detector offers many advantages including: capability to vary the target nucleus (main gas component) from He to Xe, operation at pressures from 1 Bar to 10 Bar, 4$\pi$ tracking with track thresholds down to 5 MeV, exquisite particle ID which allows for very prcise determination of exclusive final states and the addition of a magnet field allows for energy measurement via spectrometry as well as calorimetry (from the ECAL).  In DUNE, ND-GAr allows functions as muon catcher for the pixelated LAr detector which is just upstream.  The return iron has a window which allows muons that exit the LAr to be accurately momentum analyzed in the HPgTPC.  A detector such as ND-GAr provides all of the design features enumerated in Section~\ref{sec:POatND}.  Although the point resolution of a gaseous TPC does not reach the micron scale, because of the detectors excellent event reconstruction capability, kinematic analyses can be used to tag taus.

%% file: options.tex
\section{Machine options}
\label{sec:MO}

Current machine concepts for a neutrino factory are based on a proton-driven scheme as developed by the U.S. Muon Accelerator Program (MAP).  Fig.~\ref{fig:NFMCC} shows the concept for a 5~GeV muon storage ring to produce neutrinos as was considered for the NuMAX concept~\cite{Delahaye:2018yfq}. The key elements of such a machine include:
\begin{itemize}
    \item A multi-GeV, MW-class proton driver;
    \item A front end that includes proton target that produces tertiary muons, a decay channel, buncher and phase rotator;
    \item An initial 6D cooling stage capable of handling both species of muons~\cite{Alexahin}
    \item LINACs and Recirculating Linear Accelerators (RLA) to achieve the desired beam energy;
    \item A muon storage ring.
\end{itemize}
The muon energy in such a machine can be readily extended to tens of GeV by the addition of addition of one or two RLAs;  MAP designs included RLA stages to reach the 63~GeV corresponding to the beam energy for s-channel production of the Higgs boson. The muon storage ring required for these higher energies would be a straightforward extrapolation of existing muon storage ring concepts.  

The neutrino factory accelerator complex utilizes a subset of the elements needed for a high energy Muon Collider (MC), which is currently being designed by the International Muon Collider Collaboration~\cite{IMCC:web-site}.  The only accelerator complex elements that would be unique to the NF would be the transfer lines to the muon storage ring and the storage ring itself.  The early acceleration stages in the MC design would need to allow for muon extraction at the energy relevant for the muon storage ring.  While the NF would only require

\subsection{Low Energy Acceleration}
A single pass linac with a combination of 325 and 650 MHz superconducting RF (to alleviate extremely high peak-power RF) may be used to accelerate muons to 1.95\,GeV. One could configure it using  25 MV/m RF cavities with a 7.5 cm aperture radius. The initial phase-space of the beam, as delivered by the muon front-end, is characterized by significant energy spread; the linac has been designed so that it first confines the muon bunches in longitudinal phase-space, then adiabatically superimposes acceleration over the confinement motion, and finally boosts the confined bunches to 1.95 GeV. In the initial part of the linac, when the beam is still not relativistic, the far-off-crest acceleration induces rapid synchrotron motion, which allows bunch ‘head’ and ‘tail’ to switch places within the RF bucket three times during the course of acceleration. This process~\cite{Bogacz:2003uk} is essential for averaging energy spread within the bunch, which ultimately yields desired bunch compression in both bunch-length and momentum spread. The large acceptance of the linac requires large apertures and tight focusing. This, combined with moderate beam energies, favors solenoidal  rather than quadrupole focusing for the entire linac.

\begin{figure}[b!]
	       	\centering
                     \includegraphics[width=150mm]{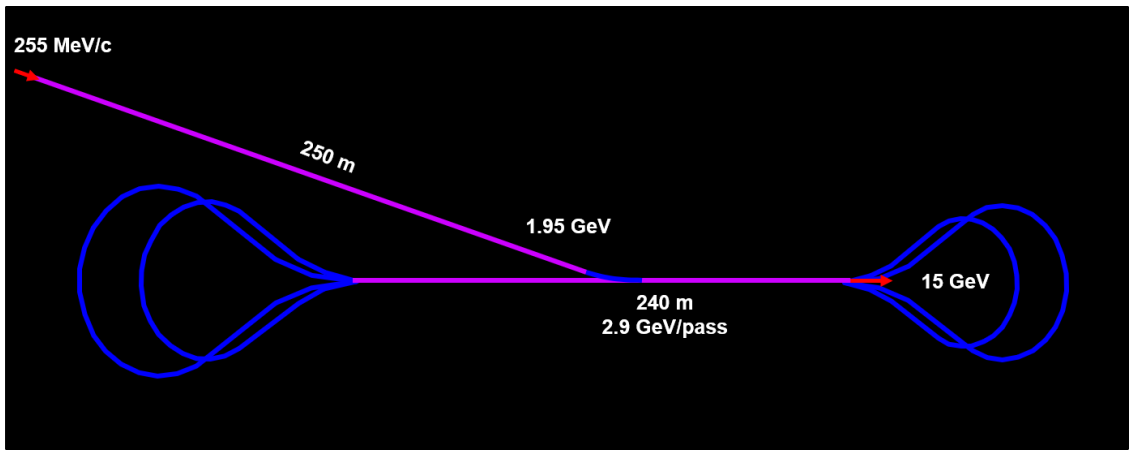}      
                     \vspace{-2mm}                             
	       	\caption{Linac plus 4.5 pass dogbone acceleration to 15 GeV.}   	
		 \label{Dogbone}  
	      \end{figure}

The linac would be followed by a Recirculating Linear Accelerator\,(RLA) configured with 2.9~GeV/pass, 650 MHz  superconducting linac based on quadrupole focusing, completed with four `droplet' arcs, where the beam reaches 15 GeV in 4.5 recirculation passes. 
This configuration is illustrated in Fig.\,\ref{Dogbone}.
The arcs use 1.6\,T dipoles.
 
The main thrust of the multi-pass RLA is its very efficient usage 
of the expensive superconducting linac. The ‘dogbone’ topology further boosts the efficiency of linac usage (close to a factor of 2 higher than a corresponding racetrack RLA), 
because  the beam is being accelerated while traversing the linac in both directions. Finally, the ‘dogbone’ topology is inherently suited for simultaneous acceleration of both $\mu^+$ and $\mu^-$ charge species; they follow the same direction through the linac, while moving in opposite directions through the `droplet' arcs. RLA acceleration from 1.95 GeV to 15 GeV further compresses and shapes the longitudinal and transverse phase-space.

\subsection{Decay Ring}

The neutrino beam is generated from stored muons decaying in flight in the decay ring.
To maximise the efficiency of the neutrino beam generation the decay ring needs to be equipped with long straight sections with relatively large $\beta$ function, which reduces the divergence of the beam, and compact arcs. Neutrino detectors can be placed along the long straight sections, either close to the ring as near detectors or far away, if the long baseline experiment is conducted. If the neutrino beam needs to be sent to the far detector, the straight section needs to be oriented with an appropriate angle dictated by the length of the baseline. Several such rings have been already designed with various geometries like a racetrack, a triangle or a bow-tie~\cite{Prior:2009zzc}. 

One of the most advanced designs to date was proposed within the IDS-NF~\cite{IDS-NF:2011swj} study, which is based on the racetrack geometry~\cite{Kelliher:2013fva}. This 1575.8\,m in circumference ring allows to simultaneously store three positive and three counter circulating negative muon bunches at 10\,GeV. The compact arcs and the large $\beta$ production straights optics are based on the FODO structures. The ring uses the separate function magnets in all different optical superperiods. The arcs are equipped with 4\,T SC dipoles, which is a compromise between the need of compactness and the cost. The injection systems for both charges are located in special insertions between the arc and the production straights and their optics is based on the triplet cells to maximise the available length for septa and kickers. The focusing in the insertions is strong enough to reduce the beam size. 

Once the maximum energy required for the neutrino beam generation is established, the design of the ring can be performed. A potential length of the long baseline will set the ring orientation. Concerning the shape of the ring, racetrack seems to be the most advantageous from the point of view of maximising the neutrino production efficiency and the cost. While the size of the ring is dictated by the maximum required energy, the strength of the bending field in the dipoles and the time structure of the beam (a bunch train length), the magnet apertures are set by the beam size at the smallest energy. Once the ring is designed for the highest energy, scaling down should be straightforward assuming the quadrupole families are independently powered to assure the control of the working point and the matching of optical functions between various modules.

%% file: synergies.tex
\section{Synergies with the Muon Collider and FNAL accelerator complex}
\label{sec:Synergies}

\begin{figure}[t]
\begin{center}
\includegraphics[width=0.9\textwidth]{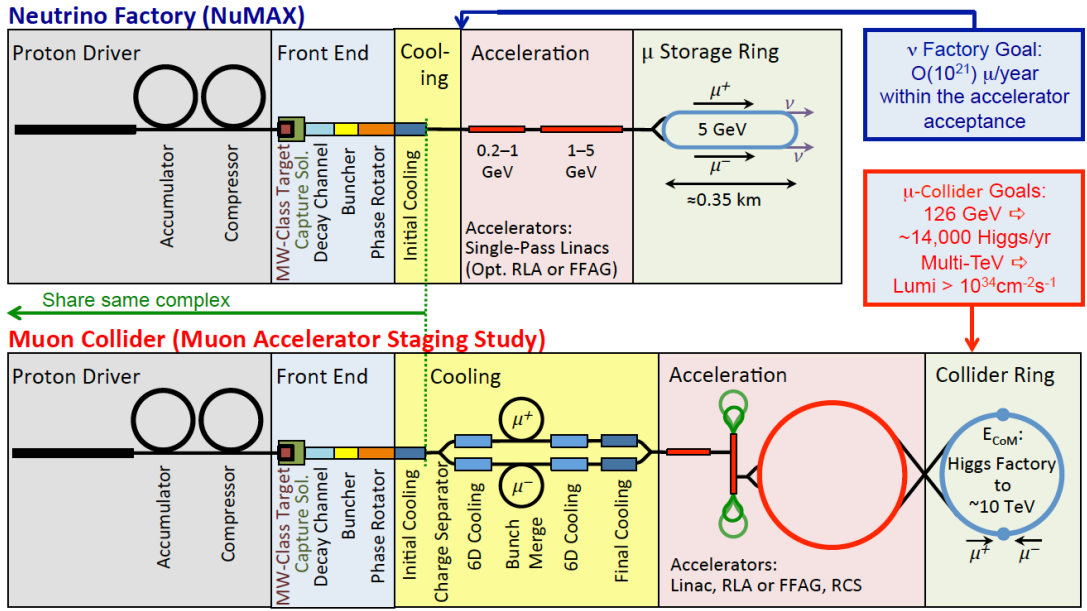}
\caption{Block diagrams showing the principal elements of a Neutrino Factory (NF) and a Muon Collider (MC).}%
\label{fig:NFMCC}
\end{center}
\end{figure}

\begin{figure}[h]
\begin{center}
\includegraphics[width=0.9\textwidth]{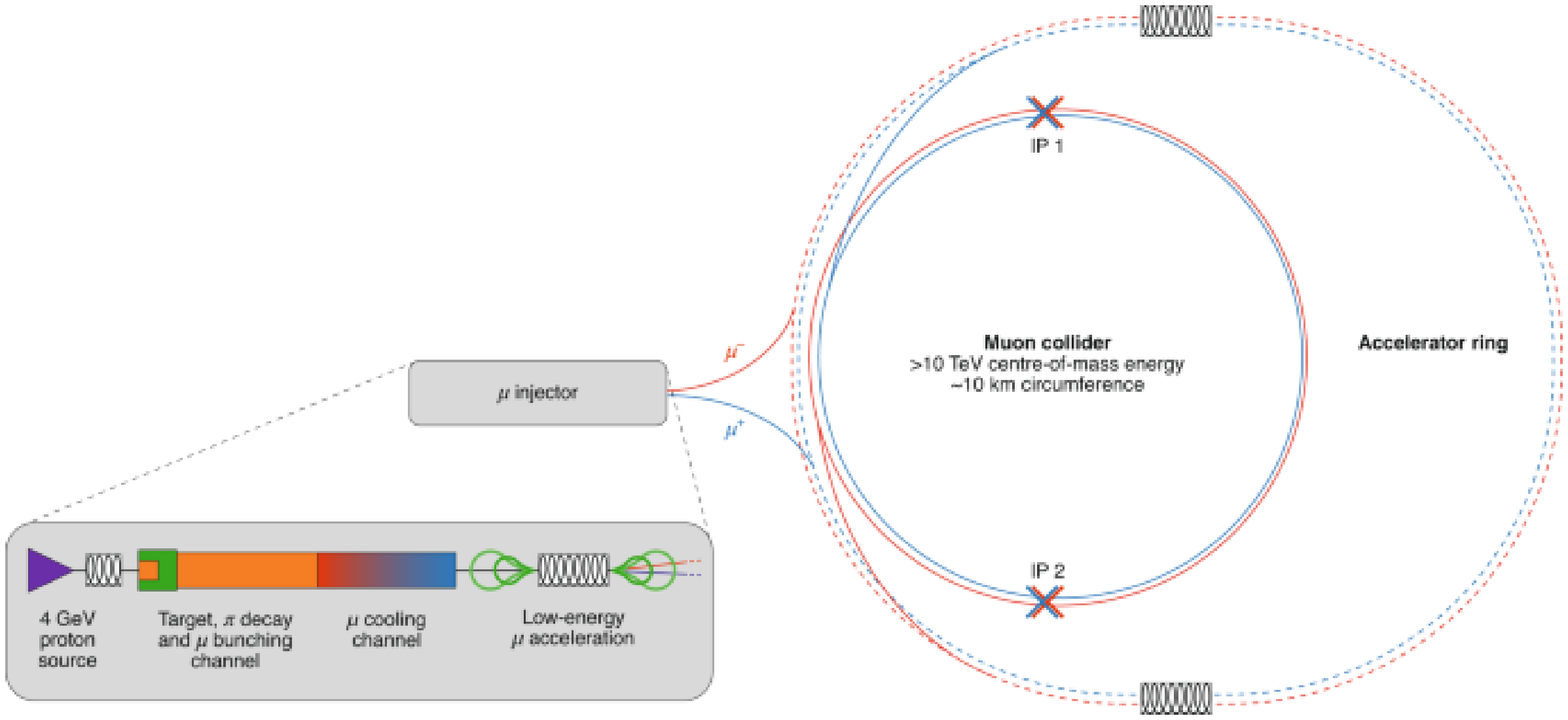}
\caption{Schematic layout of 10 TeV-class muon collider complex being study within the International Muon Collider Collaboration.  From https://muoncollider.web.cern.ch/}%
\label{fig:IMCC}
\end{center}
\end{figure}

A Fermilab Muon Collider site filler is currently under study (see ``Future Collider Options for the US'' white paper submitted to the Proceedings of the US Community Study on the Future of Particle Physics) but the concept dates back to the early 2000s.  The required parameter space towards a 10 TeV MC site filler has been identified and a first design concept has been developed, which is described in the above mentioned white paper.  The US Muon Accelerator Program (MAP)~\cite{Palmer:2013/07/02bta}, which functioned between 2011 and 2016, was tasked to assess the feasibility of the technologies required for the construction  of both the Neutrino Factory and the Muon Collider. At the conclusion of MAP, the program had produced a number of significant milestones including an End-to-End simulation of cooling for the MC~\cite{Palmer:2016gws}.  What this study showed was that the front-end for the Neutrino Factory and that for the Muon Collider can be essentially identical.  This is illustrated in Figure~\ref{fig:NFMCC}.
As can be seen from this figure, the front end (up to initial cooling) is the same for both facilities.  Although MAP was terminated in 2016, work continued on documenting the program's results and has provided a ``jumping-off'' point for the recently formed International Muon Collider Collaboration, IMCC~\cite{IMCC:web-site}.  The design strategy taken by the IMCC relies heavily on the concepts developed by the MAP collaboration. In the baseline design, muons are produced in decays from pions produced by colliding a multi-megawatt proton beam onto a target.  The IMCC envisions a staged approach for the deployment of a Muon Collider with the first stage collider operating at the center-of-mass energy of 3 TeV and the second stage at 10+ TeV, which is shown in Figure~\ref{fig:IMCC}.
The front-end labeled as $\mu$ injector in  Figure~\ref{fig:IMCC} can also function as the $\mu$ injector for a Neutrino Factory.  The Neutrino Factory could be part of a staging scheme for the IMCC, but nevertheless, the R\&D directed at the $\mu$ injector is directly applicable to that needed for the Neutrino Factory.

%% file: conclusion.tex
\section{Summary \& Outlook}
\label{sec:conclusions}

LBNF-DUNE and Tokai-to-Hyper-Kamiokande are scheduled to start taking data by the end of the decade. Both will measure muon disappearance and electron appearance at their far-detector sites, while LBNF-DUNE will also be sensitive to tau appearance. Both make use of a beam of mostly muon-type neutrinos produced in the decay of sign-selected charged pions, referred to as a super-beam. The large fluxes of pions and neutrinos will also allow precision measurements of neutrino properties and interactions, and searches for new, relatively light states including neutral heavy leptons and axion-like particles.

Neutrino factories -- neutrino beams produced in the decay of a muon or antimuon beam inside a storage ring -- yield cleaner, richer, and more flexible neutrino beams relative to super-beams. In more detail: (i) Assuming the stored muon energy and flux are well known, the daughter neutrino energy spectra and total fluxes are also well known. (ii) Muons decay into both a muon-type neutrino and an electron-type antineutrino (or vice-versa for antimuons). Hence, neutrino factories provide the unique opportunity to study, concurrently, $\nu_{\mu}$ and $\nu_{e}$ oscillations in the same experimental setup and are the only known intense source of high-energy (larger than 10~MeV) electron-type neutrinos. With a neutrino factory one has, in principle, access to $\nu_{e}\to \nu_e$ $\nu_{e}\to\nu_{\mu}$, and $\nu_{e}\to\nu_{\tau}$ oscillations. These are practically inaccessible to super-beam experiments. Finally, (iii) in a neutrino factory, one has the flexibility to choose the muon energy without significant loss of neutrino flux.      

In the context of long-baseline oscillations, the expected impact of the neutrino factory will depend strongly on the results revealed by DUNE and Hyper-K. Assuming DUNE and Hyper-K do not run into any surprises (other than, for example, establishing that CP-invariance is violated in the neutrino sector), Neutrino Factory setups are expected to measure, and in some cases over-measure, oscillation parameters with higher precision. On the other and, should Hyper-K and DUNE discover more new physics in the lepton sector, neutrino factories will, likely, have the ability to explore the new physics in more detail, perhaps addressing questions outside the reach of superbeam experiments. Finally, should new ``anomalies'' come out unresolved once data from DUNE and Hyper-K are analyzed, the different and more diverse beam conditions offered by the neutrino factory are likely to allow one a clean path for resolving these hypothetical anomalies. If the ``anomalies'' suggest a need for more detailed studies at high neutrino energy, the neutrino factory maybe the only path.  Detailed phenomenological and theoretical studies of all these hypothetical scenarios are necessary and yet to be performed.

The intensity of the neutrino factory, along with the superior-quality beam, also allow a rich near-detector program. Many concrete examples were discussed in some detail here but dedicated analyses are not yet available. The stored-muon beam also allows one to carry out a comprehensive neutrino scattering program that may prove necessary in order to reach ultimate precision and sensitivity in long-baseline oscillation experiments, independent from the nature of the long-baseline experimental setup. 

Our ability to exploit the neutrino factory beam depends on the capabilities of the detectors that sit on the other end. We reviewed different ideas here. Important questions include whether the detectors can distinguish electrons, muons, and photons, identify the charge of muon tracks and electron tracks, faithfully reconstruct the incoming neutrino energy, etc. Different detector-choices will allow access to different physics questions. 

The challenges associated with building a neutrino factory have also been discussed in some detail. Active, continuous, dedicated research and development today are necessary if the community wants to be prepared to start building a neutrino factory in the next decade. Such an effort extends beyond the ambitions of the neutrino program and is perfectly synergistic with efforts to build a muon collider, one of the identified options for both understanding the physics of the Higgs boson and exploring physics at energy scales that significantly supresede those of the LHC (at the parton level). Finally, both neutrino factories and muon colliders appear to be an excellent ``fit'' to the Fermilab accelerator complex. Both would be exquisite flagships for the laboratory in the late 2030s and beyond. 